\documentclass[10pt,a4paper]{article}

\usepackage[utf8]{inputenc}
\usepackage[T1]{fontenc}
\usepackage{geometry}
\usepackage{authblk}
\usepackage{graphicx}
\usepackage{url}
\usepackage[super,sort&compress]{natbib}
\usepackage[colorlinks=true,linkcolor=blue,citecolor=blue,urlcolor=blue]{hyperref}

\usepackage{listings}
\usepackage[table]{xcolor}
\usepackage{booktabs}
\usepackage{caption}
\usepackage{float}
\usepackage{siunitx}
\usepackage{array}
\usepackage{multirow}

\definecolor{diffadd}{RGB}{34,139,34}
\definecolor{diffremove}{RGB}{178,34,34}
\definecolor{diffctx}{RGB}{128,128,128}

\lstdefinelanguage{diff}{
  basicstyle=\ttfamily\footnotesize,
  morecomment=[f][\color{diffadd}]{+},
  morecomment=[f][\color{diffremove}]{-},
  morecomment=[f][\color{diffctx}]{@},
  keepspaces=true,
  frame=single,
  rulecolor=\color{black!30},
  breaklines=true,
  columns=fullflexible
}

\title{A Function-level Dataset of Vulnerable and Fixed\\Source Code in JavaScript and TypeScript}

\author[1,*]{Tamás Viszkok}
\author[1]{Péter Hegedűs}

\affil[1]{University of Szeged, Szeged, Hungary}

\date{}

\begin{document}

    \maketitle

    \begingroup
    \renewcommand{\thefootnote}{\fnsymbol{footnote}}
    \footnotetext[1]{Corresponding author}
    \endgroup

    {\let\thefootnote\relax\footnotetext{\textit{Email addresses: tviszkok@inf.u-szeged.hu (T. Viszkok), hpeter@inf.u-szeged.hu (P. Hegedűs)}}}

    \begin{abstract}
        JavaScript and TypeScript are widely used in modern web development, making their security critical; however, automated vulnerability detection is often constrained by the availability of high-quality training data. Here we present JsVul, a dataset curated from seven major sources. Unlike generic multi-language datasets that may retain noise -- such as minified code and cosmetic edits -- JsVul utilizes a language-specific pipeline. We collected pre-fix and post-fix versions of files around security fixes and, by filtering irrelevant artifacts and applying automated syntax normalization, isolated security-related changes. We ensured data integrity through multi-stage deduplication and heuristic-based labeling. Provided in a time-ordered JSONL format, JsVul supports robust model training in the JavaScript and TypeScript ecosystem and demonstrates the importance of language-aware preprocessing in building vulnerability datasets.
    \end{abstract}

    \section*{Background \& Summary}

    The widespread use of JavaScript in both client-side and server-side environments has made it a prime target for attackers, particularly within the npm ecosystem where high interconnectivity allows vulnerabilities to propagate rapidly\cite{zimmermann2019small}. As the software industry increasingly adopts Large Language Models (LLMs) for code generation, concerns have arisen regarding the security of the generated code, as models may reproduce vulnerabilities present in their training data\cite{pearce2022asleep}. Consequently, the development of robust machine learning-based vulnerability detection systems depends on the availability of high-quality, reliable training data\cite{guo2023investigation,croft2023data}.

    Constructing such datasets presents specific difficulties. Foundational approaches, such as DeepBugs\cite{pradel2018deepbugs} and VulDeePecker\cite{li2018vuldeepecker}, demonstrated the potential of applying probabilistic models and deep learning to vulnerability detection. However, in the broader domain of AI for Code, code duplication has been shown to artificially inflate model performance metrics, leading to poor generalization on novel data\cite{allamanis2019adverse}. Crucially, subsequent evaluations\cite{chakraborty2021deep} established that realistic class distributions and rigorous deduplication are essential for valid evaluation. Furthermore, standard experimental setups often ignore the evolution of vulnerabilities over time, introducing ``temporal bias'' (or data snooping) where models inadvertently learn from future patterns to predict past events\cite{pendlebury2019tesseract,arp2022dos}. While benchmarks like Big-Vul\cite{fan2020ac} have standardized assessment for C/C++, the JavaScript ecosystem has historically lacked a comparable resource utilizing strict cleaning methodologies.

    Vulnerability dataset curation generally follows either \textit{manual curation} or \textit{automated mining}. Manual curation yields high precision but is difficult to scale, whereas automated mining scales well but often introduces noise. Datasets providing only metadata (repository URLs and commit hashes) introduce the risk of "link rot," where data is lost when URLs become invalid due to repositories being deleted or made private. Furthermore, multi-language datasets may not account for ecosystem-specific artifacts, such as minified code or the flexible syntax of JavaScript. Without language-specific filtering, automated scrapers may include irrelevant files (e.g., tests, build configurations) or cosmetic edits, inflating the scope of modified lines and misleading automated detection methods.

    JsVul employs a hybrid approach to bridge the gap between precision and scale. We aggregated raw fixing commits from seven sources using automated retrieval, then refined these data through a pipeline that integrates language-specific automated cleaning with targeted manual verification. By applying JavaScript-specific syntax normalization and custom heuristics for minified code, JsVul isolates security-relevant fixes, resulting in a function-level dataset suitable for deep learning applications.

    \section*{Methods}

\begin{figure}[htbp]
        \centering
        \includegraphics[width=\linewidth]{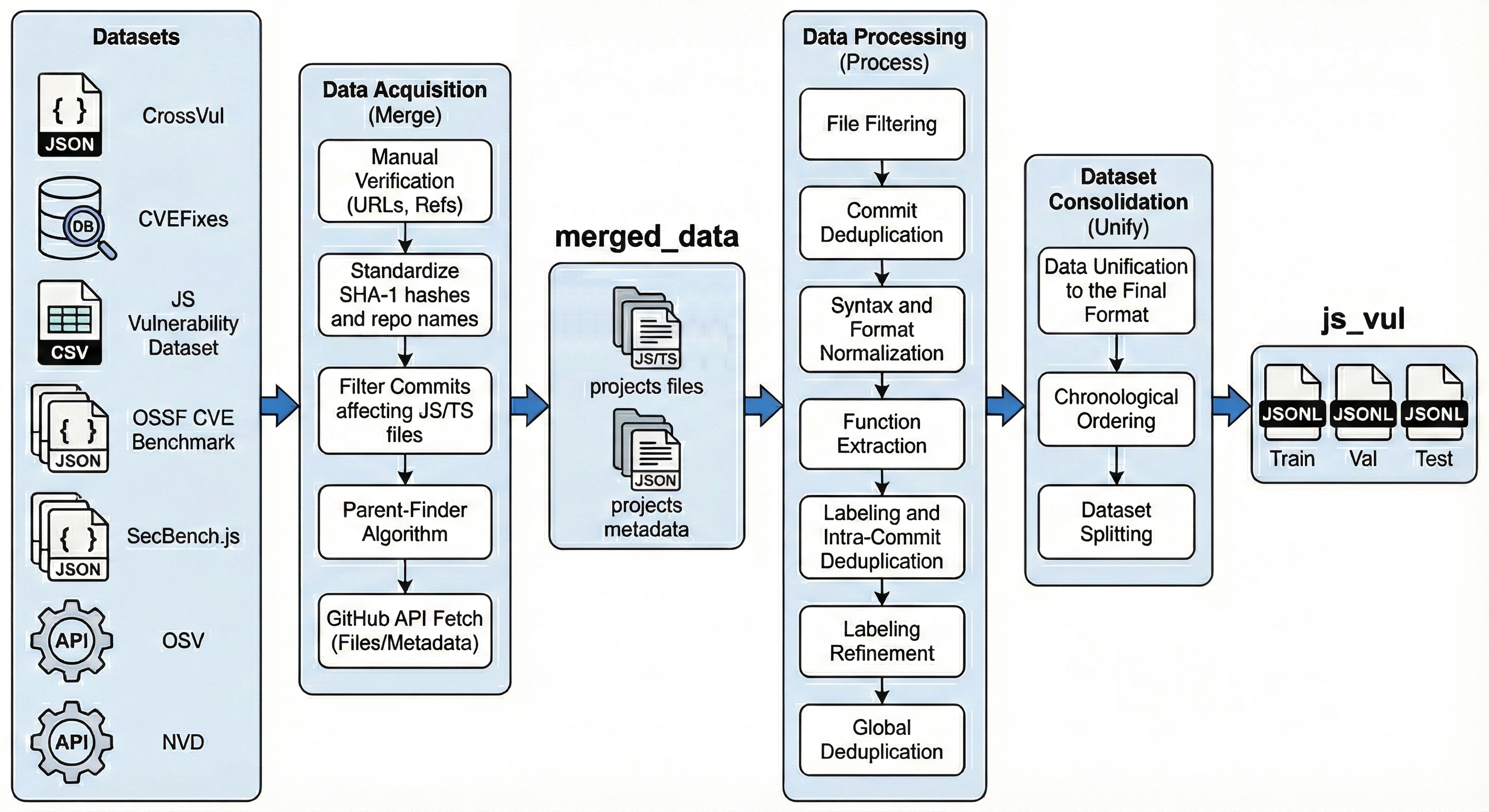} 
        \caption{Overview of the data collection and processing pipeline. The system aggregates multiple vulnerability datasets, standardizes metadata via the GitHub API, and applies strict filtering and custom labeling to ensure data quality.}
        \label{fig:pipeline_overview}
    \end{figure}

    JsVul was constructed using a three-stage pipeline -- \textit{Merge}, \textit{Process}, and \textit{Unify} -- designed to transform raw metadata into a structured dataset (see Figure~\ref{fig:pipeline_overview}). The pipeline addresses challenges including minified code, duplicated commits, syntax errors, and temporal data leakage.

    \subsection*{Data Acquisition (\textit{Merge})}
    We aggregated fixing commits from seven primary sources. For static datasets, data were downloaded directly from the referenced archives: \textbf{CrossVul}\cite{crossvul}, \textbf{CVEFixes}\cite{cvefixes}, \textbf{JS Vulnerability Dataset}\cite{jsvuln}, \textbf{OSSF CVE benchmark}\cite{ossfcvebenchmark}, and \textbf{SecBench.JS}\cite{secbenchjs}.

    For continuously updated sources, we recorded specific retrieval dates to ensure reproducibility:
    \begin{itemize}
        \item \textbf{OSV}\cite{osv}: Retrieved on 2025-11-18 15:20 CET directly from the Google Cloud Storage bucket.
        \item \textbf{NVD}\cite{nvd}: Collected on 2025-10-31 15:23 CET via our pipeline's collection module, which queries the NVD API.
    \end{itemize}

    We encountered significant data quality issues impacting the identification, selection, and retrieval of relevant commits. First, \textbf{metadata validity} was often compromised; for instance, the OSSF CVE Benchmark\cite{ossfcvebenchmark} in \texttt{CVE-2020-11021.json} mentioned a commit in its ``prePatch'' field that modified only test files, while the ``postPatch'' commit in \texttt{CVE-2018-8035.json} does not exist. Similarly, SecBench.js\cite{secbenchjs} occasionally referenced commits updating only non-JavaScript files (e.g., citing the C++ related commit \texttt{fe52854} in \texttt{incubator/hermes-engine\_0.6.0/package.json}) or test-only commits rather than the actual fix (e.g., citing \texttt{b80d699} instead of \texttt{9c74056} in \texttt{redos/html-dom-parser\_0.1.2/package.json}), while the JS Vulnerability Dataset\cite{jsvuln} sometimes identified a ``vulnerable'' parent commit several versions older than the direct parent (e.g., citing \texttt{83cd07f} instead of \texttt{40d73e2} from project \texttt{nodebb/nodebb}), unnecessarily inflating the diff. Second, \textbf{resource identifiers} in continuously updated sources like OSV\cite{osv} occasionally contained malformed URLs (e.g., \texttt{...github.com/tensorflow/issues/...} instead of \texttt{...github.com/tensorflow/tensorflow/issues/...} in \texttt{CVE-2021-29617.json}). Third, \textbf{repository evolution} posed a universal challenge; as repositories are frequently renamed or transferred over time, datasets often contained the same fixing commit URL but with different repository names, which leads to duplication if these divergent references are not unified. Fourth, \textbf{inconsistent commit hash formatting} led to duplication; for example, in the OSSF CVE Benchmark, the ``postPatch'' hash in \texttt{CVE-2017-18352.json} differed from \texttt{CVE-2017-18353.json} only by missing characters. Fifth, \textbf{content integrity and storage} methods required overhaul; CrossVul\cite{crossvul} entries occasionally contained HTTP ``404 Not Found'' error messages saved as file content, while CVEFixes\cite{cvefixes} utilized a monolithic SQLite database that created performance bottlenecks. Finally, the \texttt{programming\_language} field in the \texttt{file\_change} table of CVEFixes was often unreliable, classifying files with C++, Java or Go extensions as JavaScript, and vice versa.
    
    To resolve these, we first manually verified and corrected erroneous commit references and URLs encountered during the merge process. Subsequently, to address the broader inconsistencies, our pipeline treats the source datasets primarily as seed lists of fixing commits; while we do not rely on their provided metadata (except the fixing commit references) or file contents for processing, we preserve the original identifiers and descriptive attributes to ensure traceability and completeness. We engineered the pipeline to query the GitHub API for every entry, utilizing the authoritative response to standardize commit hashes to full-length SHA-1 identifiers, resolve up-to-date repository names, and strictly retain only those fixing commits that modify at least one JavaScript or TypeScript file. Furthermore, to address the frequent absence or inaccuracy of pre-fix commit references in the source data, we implemented an algorithm to identify the vulnerable parent commit (see \texttt{get\_parent\_sha} in \texttt{merge\_datasets/merge.py}\cite{jsvul}), which selects the direct parent for single-parent commits and resolves the appropriate parent in the case of merge commits to ensure a reliable pre-fix state. Using these validated parameters, the pipeline retrieves the fixing commit's metadata alongside the fresh source files for both the post-fix and identified pre-fix versions -- preserving their original relative file paths -- and unifies the standardized metadata into the format described as \texttt{merged\_data} in the \hyperref[sec:datarecords]{Data Records} section.

    \subsection*{Data Processing (\textit{Process})}

    This stage addresses noise reduction and labeling accuracy. Among the sourced datasets, only the JS Vulnerability Dataset\cite{jsvuln}, OSSF CVE Benchmark\cite{ossfcvebenchmark}, and SecBench.js\cite{secbenchjs} provided function- or line-level vulnerability labels, yet each required refinement. We observed that the JS Vulnerability Dataset included test files (e.g., \texttt{spec/quantitiesSpec.js}) contrary to the filtering described in their publication, and potential flaws in their automated processing often mislabeled code units (e.g., flagging unmodified functions in \texttt{actionhero/initializers/utils.js} as vulnerable). Similarly, the OSSF CVE Benchmark occasionally implicated files unrelated to the fix; for instance, in \texttt{CVE-2017-16018.json}, \texttt{lib/server.js} is marked as the vulnerability source despite being unrelated to the fix and remaining unchanged by the fixing commit \texttt{24c57ce}. SecBench.js also exhibited data gaps, with missing line-level information in entries such as \texttt{incubator/command\_injection/heroku-addonpool\_0.0.1/package.json}. To overcome these inconsistencies and the absence of granular labels in other sources, our processing pipeline disregards the provided vulnerability metadata. Instead, it employs custom labeling heuristics along with specific filters, normalizers, and extractors to independently isolate vulnerable functions:

    \begin{enumerate}
        \item \textbf{File Filtering}: Non-JS/TS files, test files, and build/configuration files were removed. To reduce duplication, minified files—which cannot always be identified by filename alone—were filtered using a heuristic-based detection script. Each file was preprocessed with our custom utility, \texttt{js-minify-helper}, to remove comments and replace regex literals with placeholders. The script evaluated the file based on average line length, ratio of long lines, whitespace density, punctuation frequency, and single-letter identifier prevalence.

        \item \textbf{Commit Deduplication}: Duplicated commits were identified by checking for intersections in the lists of file hashes corresponding to the files modified by each entry. Overlapping entries were manually verified, retaining the version that best isolated the fix.

        \item \textbf{Syntax and Format Normalization}: To ensure diffs reflect only semantic changes, code structure was normalized. We applied \textbf{ESLint}\cite{eslint} to both pre- and post-fix files to standardize variable declarations and keyword usage (e.g., \textit{let} vs \textit{const}), followed by \textbf{Prettier}\cite{prettier} to standardize formatting. For an illustrative example, see Supplementary Figure S1 and S2. All diffs and file hashes were then regenerated.

        \item \textbf{Function Extraction}: Using our custom tool, \texttt{js-function-extractor}, we generated Abstract Syntax Trees (AST) via Babel\cite{babel}, then utilized \texttt{recast}\cite{recast} to extract and reprint function bodies, stripping comments to isolate executable logic.

        \item \textbf{Labeling and Intra-Commit Deduplication}: Initially, functions were labeled as "vulnerable" if they appeared in the pre-fix file and intersected with lines modified by the fixing commit. However, this line-based approach misclassifies moved functions or those with cosmetic edits as modifications. To resolve this, we performed deduplication within single commits. We leveraged the fact that moved functions are inherently identical in content, while our normalization process ensures that functions with only cosmetic edits also become identical to their post-fix counterparts. By filtering based on content identity, we identified these non-logic changes; where a "modified" function was found to be identical to its fixed version, we re-labeled it as non-vulnerable (see Figure~\ref{fig:normalization_example} for an example).

        \item \textbf{Labeling Refinement}: After isolating semantically modified functions, we applied filtering heuristics proposed by previous research\cite{primevul}:
        \begin{itemize}
            \item \textit{onefunc:} Retains commits where exactly one function is identified as vulnerable.
            \item \textit{nvdcheck:} Retains functions explicitly mentioned in the NVD entry, or cases where a file was mentioned and contained only one vulnerable function.
        \end{itemize}
        Commits failing both checks were discarded.

        \item \textbf{Global Deduplication}: Finally, functions were deduplicated across the entire dataset. If identical functions existed with the same label, the chronologically first entry based on \texttt{publish\_time} was retained. If identical functions had conflicting labels, we prioritized the earliest instance labeled as "vulnerable".
    \end{enumerate}

    \begin{figure}[htbp]
        \centering
        \begin{minipage}{0.95\linewidth}
            \begin{lstlisting}[language=diff]
@@ -41,7 +42,10 @@ function readFile(path) {
 
 async function restore(file, refLog) {
   refLog.logData.push({ color: "lawngreen", Message: "Starting Restore" });
-  refLog.logData.push({ color: "yellow", Message: "Restoring from Backup: " + file });
+  refLog.logData.push({
+    color: "yellow",
+    Message: "Restoring from Backup: " + file,
+  });
   const pool = new Pool({
     user: postgresUser,
     password: postgresPassword,\end{lstlisting}
        \end{minipage}
        \caption{\textbf{An example of a cosmetic edit filtered by Intra-Commit Deduplication.} Both formats are valid under Prettier, but our pipeline's intra-commit deduplication step with the underlying function extractor correctly filters this change to avoid a false positive.}
        \label{fig:normalization_example}
    \end{figure}

    \subsection*{Dataset Consolidation (\textit{Unify})}
    In the final stage, data were structured into JSON Lines (JSONL) format. Each entry contains a unique ID, the function \texttt{body}, a binary \texttt{label}, and metadata as described in the \hyperref[sec:datarecords]{Data Records} section. As noted by Pendlebury et al.\cite{pendlebury2019tesseract} and Arp et al.\cite{arp2022dos}, random splitting strategies (e.g., $k$-fold cross-validation) introduce temporal bias in security datasets. To mitigate this, we strictly order the dataset by \texttt{publish\_time} and employ a chronological split, dividing the data into training (80\%), validation (10\%), and test (10\%) sets.

    \section*{Data Records}
    \label{sec:datarecords}

    The data generated in this study are available on Zenodo\cite{dataset_citation}. The repository contains three compressed archives.

    \subsection*{Archives \texttt{js\_vul} and \texttt{js\_vul\_pairs\_only}}
    These archives contain the final JsVul dataset in JSONL format. The \texttt{js\_vul} archive contains the complete dataset, while \texttt{js\_vul\_pairs\_only} contains only paired examples where both the vulnerable (pre-fix) and fixed (post-fix) versions exist.

    Each JSON object adheres to the following schema:

    \begin{enumerate}
        \item \textbf{id} (string): Unique identifier formatted as: \\ \texttt{"[github\_repo]::[commit\_hash]::[file\_path]::[start\_line]::[start\_column]"}.
        \item \textbf{paired\_id} (string, optional): The unique identifier of the corresponding vulnerable or fixed counterpart, if applicable.
        \item \textbf{project}, \textbf{sha}, \textbf{file} (string): Repository name, commit hash, and relative file path.
        \item \textbf{loc} (json object): Function location coordinates in the formatted files: \textbf{start\_line}, \textbf{start\_column}, \textbf{end\_line}, and \textbf{end\_column}.
        \item \textbf{body} (string): The source code of the function.
        \item \textbf{label} (int): The vulnerability label of the function (1 for vulnerable, 0 for non-vulnerable).
        \item \textbf{name} (string, optional): The function name.
        \item \textbf{cwe} (string array, optional): Common Weakness Enumeration (CWE) identifiers.
        \item \textbf{cve}, \textbf{ghsa}, \textbf{snyk}, \textbf{other} (string array, optional): Vulnerability identifiers from NVD, GitHub Security Advisories (GHSA), Snyk, or other sources.
        \item \textbf{publish\_time} (json object, optional): Publication date (\textbf{year}, \textbf{month}, \textbf{day}).
    \end{enumerate}

    \subsection*{Archive \texttt{merged\_data}}
    This archive contains the unified, pre-processed dataset resulting from the Merge phase, structured in two subdirectories:

    \begin{itemize}
        \item \texttt{files}: Contains source code files (pre- and post-commit) organized by repository and commit SHA.
        \item \texttt{metadata}: Contains JSON files for each project, where keys are fixing commit hashes. Metadata fields include: \textbf{cwe}, \textbf{cve}, \textbf{github}, \textbf{snyk}, \textbf{commit\_msg}, \textbf{additions}, \textbf{deletions}, \textbf{changes}, \textbf{files}, \textbf{vuln\_sha}, and \textbf{sources}.
    \end{itemize}

    \section*{Technical Validation}

    To ensure the fidelity of the JsVul dataset, we performed validation focusing on data purity, redundancy removal, and semantic relevance. Table~\ref{tab:filtering} summarizes the impact of each pipeline stage.

    \subsection*{Noise Reduction via Syntax Normalization}
    We compared diff sizes and the number of affected functions before and after applying ESLint and Prettier. The normalization process resulted in a net reduction in the average number of lines modified per commit, confirming that a portion of the original commit data consisted of non-semantic formatting noise. To ensure integrity, we verified that the number of affected functions after formatting was less than or equal to the count before formatting. We manually inspected 25\% of randomly selected changes to ensure the code was not modified semantically.

    \subsection*{Deduplication and Integrity Checks}
    \begin{itemize}
        \item \textbf{Dataset-wise Commit Deduplication:} We manually verified all 97 commit pairs flagged as duplicates. We resolved 87 cases by merging metadata into a single entry, unifying instances where different sources referenced the same issue with different hashes, while the rest were kept as is.
        \item \textbf{Commit-wise Function Deduplication:} We identified duplicate functions with different labels within commits. Those with no semantic changes were relabeled as non-vulnerable. Filtering non-semantic changes not only improved quality by eliminating false positives but also slightly increased the dataset size, as it allowed more commits to meet the single-function requirement of \texttt{onefunc} and \texttt{nvdcheck}. Manual comparison revealed that among vulnerable functions, 12 instances were removed (false positives) and 17 new instances were recovered.
        \item \textbf{Dataset-wise Function Deduplication:} Conflicts where the same function appeared as both "vulnerable" and "fixed" across different commits were resolved by retaining the "vulnerable" label to minimize false negatives.
    \end{itemize}

    \begin{table}[ht]
        \centering
        \caption{\textbf{Technical validation of the curation pipeline.} Metrics demonstrating the reduction of noise and preservation of relevant code units through each processing stage.}
        \label{tab:filtering}
        \vspace{0.2cm}
        \resizebox{\linewidth}{!}{%
        \begin{tabular}{|l|c|c|c|c|c|}
            \hline
            \textbf{Stage} & \textbf{Commits} & \textbf{Files} & \textbf{Changed lines} & \textbf{Affected functions} & \textbf{Vulnerable / Non-vulnerable} \\
            \hline
            merged        & 3,523           & 22,359         & 1,521,272             & -                          & - \\
            filtered      & 3,310           &  7,557         &   192,947             & -                          & - \\
            dedup\_c      & 3,223           &  7,380         &   187,943             & 22,053                     &         10,440 / 157,111 \\
            normal        & 3,218           &  7,314         &   162,276             & 20,437                     &          9,632 / 157,051 \\
            dedup\_f      & 1,967           &  2,001         &    48,302             &  4,607                     &          2,079 /  26,190 \\
            final         & 1,960           &  1,994         &    48,018             &  4,391                     & \textbf{2,039} / \textbf{19,862} \\
            final\_po     & -               & -              & -                     & -                          & \textbf{1,342} / \textbf{1,342} \\
            \hline
        \end{tabular}%
        }
        \vspace{0.1cm}
        \raggedright
        \footnotesize
        \textbf{Row Definitions:} \textit{merged}: combined raw data; \textit{filtered}: exclusion of test/config/minified files; \textit{dedup\_c}: dataset-wise commit deduplication; \textit{normal}: syntax normalization; \textit{dedup\_f}: commit-wise function deduplication; \textit{final}: dataset-wise function deduplication; \textit{final\_po}: subset containing only matched vulnerable/fixed pairs.
    \end{table}

    \subsection*{Semantic Signal Verification}

    To verify that the dataset contains distinct semantic signals distinguishable by modern architectures (and is not merely noise), we conducted a technical validation using three distinct architectures on the \texttt{js\_vul\_pairs\_only} subset: \textbf{CodeBERT}\cite{codebert} (encoder-only), \textbf{Qwen2.5-Coder-7B-Instruct}\cite{qwencoder} (open-weight coding LLM), and \textbf{Gemini 3 Pro} (closed-source LLM, zero-shot). As shown in Table~\ref{tab:baselines}, the progression in scores confirms that the dataset presents a consistent, learnable signal.

    \begin{table}[ht]
        \centering
        \caption{\textbf{Baseline verification on the JsVul test set (pairs subset).} These metrics serve as a technical validation of the dataset's learnability and utility across different model architectures. The $F_{0.5}$-score is reported to emphasize precision.}
        \label{tab:baselines}
        \vspace{0.2cm}
        \begin{tabular}{|l|c|c|c|c|c|}
            \hline
            \textbf{Model Name} & \textbf{Method} & \textbf{Accuracy} & \textbf{Precision} & \textbf{Recall} & \textbf{F$_{0.5}$-Score} \\
            \hline
            CodeBERT & Fine-tuned & 0.5587 & 0.5032 & \textbf{1.0000} & 0.5063 \\
            Qwen2.5-Coder-7B-Instruct & Fine-tuned & \textbf{0.6030} & \textbf{0.6138} & 0.5633 & \textbf{0.6044} \\
            Gemini 3 Pro & Zero-shot & 0.5696 & 0.5444 & 0.8544 & 0.5870 \\
            \hline
        \end{tabular}

        \vspace{0.1cm}
        \footnotesize
        Experimental parameters are listed in Supplementary Table S1.
    \end{table}

    \section*{Usage Notes}
    The JsVul dataset is distributed in JSONL format to allow for efficient line-by-line processing.

    \subsection*{Data Structure}
    Each record contains the extracted \texttt{body} of the function and a binary \texttt{label} (1 for vulnerable, 0 for non-vulnerable); for an illustrative example, see Supplementary Figure S3.

    \subsection*{Temporal Splitting}
    To prevent temporal bias\cite{pendlebury2019tesseract}, the dataset is ordered chronologically by \texttt{publish\_time}. We recommend adhering to the provided temporal splits (80\% training, 10\% validation, 10\% testing) to accurately simulate vulnerability forecasting.

    \section*{Funding}
    This work was supported by the Ministry of Culture and Innovation of Hungary from the National Research, Development and Innovation Fund, through the K\_23, OTKA Funding Scheme, under Project K 147225.
    The publication charge was covered by the University of Szeged Open Access Fund (Grant Nr. 8358).

    \section*{Author Contributions}
    T.V. designed the study, implemented the pipeline, and drafted the manuscript. P.H. supervised the project and reviewed the manuscript.

    \section*{Competing Interests}
    The authors declare no competing interests.

    \section*{Data Availability}
    The data generated and analyzed during this study are available in the Zenodo repository\cite{dataset_citation}.

    This record contains the three compressed archives described in the Data Records section:
    \begin{itemize}
        \item \textbf{\texttt{js\_vul}}: The complete curated dataset in JSONL format.
        \item \textbf{\texttt{js\_vul\_pairs\_only}}: A subset containing only matched pairs of vulnerable and fixed functions.
        \item \textbf{\texttt{merged\_data}}: The unified pre-processed source data and metadata, including the \texttt{files} directory (source code pre- and post-commit) and \texttt{metadata} directory (JSON metadata for each project).
    \end{itemize}

    \section*{Code Availability}
    The code used to process the data, including the pipeline for filtering, normalization, and deduplication, is available at \url{https://github.com/jsvul/jsvul}. The repository includes instructions for reproducing the dataset or generating custom variants.

    \bibliographystyle{naturemag}

\clearpage
\newgeometry{margin=0.9in}
\linespread{1}\selectfont

\renewcommand{\thetable}{S\arabic{table}}
\renewcommand{\thesection}{S\arabic{section}}
\renewcommand{\thefigure}{S\arabic{figure}}
\renewcommand{\lstlistingname}{Listing}
\renewcommand{\thelstlisting}{S\arabic{lstlisting}}

\setcounter{table}{0}
\setcounter{section}{0}
\setcounter{figure}{0}
\setcounter{lstlisting}{0}

\lstdefinelanguage{diff}{
  basicstyle=\ttfamily\scriptsize,
  morecomment=[f][\color{diffadd}]{+},
  morecomment=[f][\color{diffremove}]{-},
  morecomment=[f][\color{diffctx}]{@},
  keepspaces=true,
  frame=single,
  rulecolor=\color{black!30},
  breaklines=true,
  columns=fullflexible
}

\lstset{
    basicstyle=\ttfamily\small,
    breaklines=true,
    frame=single,
    backgroundcolor=\color{gray!10},
    captionpos=b,
    columns=fullflexible,
    keepspaces=true,
    showstringspaces=false
}

\newcolumntype{L}[1]{>{\raggedright\arraybackslash}p{#1}}

\begin{center}
    {\Large\bfseries Supplementary Information for:\par}
    \vspace{0.5em}
    {\large A Function-level Dataset of Vulnerable and Fixed\\Source Code in JavaScript and TypeScript\par}
\end{center}

\vspace{1em}
\noindent\rule{\textwidth}{0.4pt}
\tableofcontents
\noindent\rule{\textwidth}{0.4pt}
\newpage

\section{Supplementary Material}

    \subsection*{Model Configurations}
	
	\begin{table}[ht]
		\centering
		\small
		\renewcommand{\arraystretch}{1.3}
		\caption{\textbf{Model configurations and hyperparameters.} See \textbf{Supplementary Listings \ref{lst:sys_prompt} and \ref{lst:user_prompt}} below for the exact prompt templates used for both Qwen and Gemini.}
		\label{tab:model_params}
		
		\addcontentsline{toc}{subsection}{Table S\arabic{table}: Model configurations and hyperparameters}
		
		\begin{tabular}{L{3.5cm} L{9.9cm}}
			\toprule
			\textbf{Parameter} & \textbf{Value / Configuration} \\
			\midrule

			\multicolumn{2}{l}{\textbf{Panel A: CodeBERT}} \\
			\midrule
			Learning Rate & $4.64 \times 10^{-5}$ \\
			Weight Decay & $0.206$ \\
			Batch Size & Train: 16, Eval: 32 \\
			Optimization & F-measure ($\beta=0.5$) \\
			Checkpointing & Best model loaded at end of training \\
			\midrule

			\multicolumn{2}{l}{\textbf{Panel B: Qwen2.5-Coder-7B-Instruct}} \\
			\midrule
			\textbf{Prompts} & \textbf{See Supplementary Listings \ref{lst:sys_prompt} \& \ref{lst:user_prompt}} \\
			\addlinespace
			\textbf{Quantization (BnB)} & \\
			Config & 4-bit (nf4), Double Quant, bfloat16 \\
			\addlinespace
			\textbf{LoRA Config} & \\
			Rank / Alpha & $r=16$, $\alpha=32$ \\
			Dropout & 0.05 \\
			Target Modules & \texttt{q\_proj, k\_proj, v\_proj, o\_proj, gate\_proj, up\_proj, down\_proj} \\
			\midrule

			\multicolumn{2}{l}{\textbf{Panel C: Gemini 3 Pro}} \\
			\midrule
			\textbf{Prompts} & Text content is identical to Qwen (Listings \ref{lst:sys_prompt} \& \ref{lst:user_prompt}), but \textbf{concatenated} into a single string and passed as the prompt. \\
			Inference Settings & None set manually (Default API configuration) \\
			\bottomrule
		\end{tabular}
	\end{table}

    \newpage
    
	\subsection*{LLM Prompts}
	
	\noindent
	\begin{minipage}{\textwidth} 
		\centering
		\captionof{lstlisting}{\textbf{System Prompt used for inference.} This static instruction was provided to the model context before the user message.}
		\label{lst:sys_prompt}
		\addcontentsline{toc}{subsection}{Listing S\arabic{lstlisting}: System Prompt used for inference}
		
		\begin{lstlisting}
You are a senior cybersecurity expert specializing in JavaScript and TypeScript.
Analyze the following code snippet for security vulnerabilities (e.g., XSS, Injection,
RCE, Prototype Pollution).

Your task:
1. Determine if the code contains a security vulnerability.
2. Output ONLY ONE WORD: "VULNERABLE" if it is vulnerable, or "SAFE" if it is not.
Do not output any markdown, code blocks, or extra text. Just that one word.\end{lstlisting}
	\end{minipage}

    \vspace{1cm}

	\noindent
	\begin{minipage}{\textwidth} 
		\centering
		\captionof{lstlisting}{\textbf{User Message Construction Template.} Variable \texttt{code} represents the truncated source code snippet inserted dynamically.}
		\label{lst:user_prompt}
		\addcontentsline{toc}{subsection}{Listing S\arabic{lstlisting}: User Message Construction Template}
		
		\begin{lstlisting}
Here is the source code to analyze:

```javascript
{code}
```

INSTRUCTIONS:
1. Analyze the code above for security vulnerabilities.
2. Respond with ONLY ONE WORD: "VULNERABLE" or "SAFE".
3. Do NOT provide explanations or any extra text.\end{lstlisting}
	\end{minipage}

    \newpage

    \subsection*{Code Examples}
    \begin{figure}[H]
        \centering
        \begin{minipage}{\linewidth}
            \textbf{A. Raw Diff (Before Normalization)}
            \begin{lstlisting}[language=diff]
@@ -23,45 +23,37 @@ const NOT_FOUNT_INDEX = -1;
 const INDEX_PAGE = 'index.html';
 
 module.exports = function* (next) {
-
     const directory = config.get(configKey.DIRECTORY);
 
     // decode for chinese character
-    let requestPath = decodeURIComponent(this.request.path);
-    let fullRequestPath = path.join(directory, requestPath);
-    let stat = yield getFileStat(fullRequestPath);
+    const requestPath = decodeURIComponent(this.request.path);
+    const fullRequestPath = path.join(directory, requestPath);
+    // fix security issue
+    if (!fullRequestPath.startsWith(directory)) {
+        return yield next;
+    }
+    const stat = yield getFileStat(fullRequestPath);
 
     if (stat.isDirectory()) {
-
-        let files = yield readFolder(fullRequestPath);
+        const files = yield readFolder(fullRequestPath);
 
         if (files.indexOf(INDEX_PAGE) !== NOT_FOUNT_INDEX) {
-
             this.redirect(path.join(requestPath, INDEX_PAGE), '/');
-
         } else {
-
             this.body = buildFileBrowser(files, requestPath, directory);
             this.type = mime.lookup(INDEX_PAGE);
-
         }
-
     } else if (stat.isFile()) {
-
         this.body = yield readFile(fullRequestPath);
         let type = mime.lookup(fullRequestPath);
 
         if (path.extname(fullRequestPath) === '') {
-
             type = FALLBACK_CONTENT_TYPE;
-
         }
 
         this.type = type;
         log.verbose(logPrefix.RESPONSE, this.request.method, requestPath, 'as', type);
-
     }
 
     yield next;
-
 };\end{lstlisting}
        \end{minipage}

        \vspace{0.5cm}

        \begin{minipage}{\linewidth}
            \textbf{B. Normalized Diff (After Formatting)}
            \begin{lstlisting}[language=diff]
@@ -28,6 +28,10 @@ module.exports = function* (next) {
   // decode for chinese character
   const requestPath = decodeURIComponent(this.request.path);
   const fullRequestPath = path.join(directory, requestPath);
+  // fix security issue
+  if (!fullRequestPath.startsWith(directory)) {
+    return yield next;
+  }
   const stat = yield getFileStat(fullRequestPath);
 
   if (stat.isDirectory()) {\end{lstlisting}
        \end{minipage}
    
        \caption{\textbf{ESLint \& Prettier formatting effects.} \\
        In (A), non-functional changes like whitespace formatting and keyword updates (\texttt{let} to \texttt{const}) obscure the actual fix.
        In (B), after normalization, the cosmetic edits are absent, leaving the actual security fix clearly identifiable.
        File: \texttt{middleware/file-explorer.js} in repo \texttt{vivaxy/here} at commit \texttt{298dbab}.}
        \label{fig:normalization_example_2}
        \addcontentsline{toc}{subsection}{Figure S\arabic{figure}: Effects of formatting both pre- and post-fix versions with ESLint \& Prettier}
    \end{figure}

    \begin{figure}[htbp]
        \centering
        \begin{minipage}{\linewidth}
            \textbf{Raw Diff (Before Normalization)}
            \begin{lstlisting}[language=diff]
@@ -76,13 +76,20 @@ function parsePath(path) {
   var str = path.replace(/([^\\])\[/g, '$1.[');
   var parts = str.match(/(\\\.|[^.]+?)+/g);
   return parts.map(function mapMatches(value) {
+    if (
+      value === 'constructor' ||
+      value === '__proto__' ||
+      value === 'prototype'
+    ) {
+      return {};
+    }
     var regexp = /^\[(\d+)\]$/;
     var mArr = regexp.exec(value);
     var parsed = null;
     if (mArr) {
       parsed = { i: parseFloat(mArr[1]) };
     } else {
-      parsed = { p: value.replace(/\\([.\[\]])/g, '$1') };
+      parsed = { p: value.replace(/\\([.[\]])/g, '$1') };
     }
 
     return parsed;
@@ -107,7 +114,7 @@ function parsePath(path) {
 function internalGetPathValue(obj, parsed, pathDepth) {
   var temporaryValue = obj;
   var res = null;
-  pathDepth = (typeof pathDepth === 'undefined' ? parsed.length : pathDepth);
+  pathDepth = typeof pathDepth === 'undefined' ? parsed.length : pathDepth;
 
   for (var i = 0; i < pathDepth; i++) {
     var part = parsed[i];
@@ -118,7 +125,7 @@ function internalGetPathValue(obj, parsed, pathDepth) {
         temporaryValue = temporaryValue[part.p];
       }
 
-      if (i === (pathDepth - 1)) {
+      if (i === pathDepth - 1) {
         res = temporaryValue;
       }
     }
@@ -152,7 +159,7 @@ function internalSetPathValue(obj, val, parsed) {
     part = parsed[i];
 
     // If it's the last part of the path, we set the 'propName' value with the property name
-    if (i === (pathDepth - 1)) {
+    if (i === pathDepth - 1) {
       propName = typeof part.p === 'undefined' ? part.i : part.p;
       // Now we set the property with the name held by 'propName' on object with the desired val
       tempObj[propName] = val;
@@ -199,7 +206,10 @@ function getPathInfo(obj, path) {
   var parsed = parsePath(path);
   var last = parsed[parsed.length - 1];
   var info = {
-    parent: parsed.length > 1 ? internalGetPathValue(obj, parsed, parsed.length - 1) : obj,
+    parent:
+      parsed.length > 1 ?
+        internalGetPathValue(obj, parsed, parsed.length - 1) :
+        obj,
     name: last.p || last.i,
     value: internalGetPathValue(obj, parsed),
   };\end{lstlisting}
        \end{minipage}
    
        \vspace{0.5cm}
    
        \begin{minipage}{\linewidth}
            \textbf{Normalized Diff (After Formatting)}
            \begin{lstlisting}[language=diff]
@@ -76,13 +76,16 @@ function parsePath(path) {
   const str = path.replace(/([^\\])\[/g, "$1.[");
   const parts = str.match(/(\\\.|[^.]+?)+/g);
   return parts.map(function mapMatches(value) {
+    if (value === "constructor" || value === "__proto__" || value === "prototype") {
+      return {};
+    }
     const regexp = /^\[(\d+)\]$/;
     const mArr = regexp.exec(value);
     let parsed = null;
     if (mArr) {
       parsed = { i: parseFloat(mArr[1]) };
     } else {
-      parsed = { p: value.replace(/\\([.\[\]])/g, "$1") };
+      parsed = { p: value.replace(/\\([.[\]])/g, "$1") };
     }
 
     return parsed;\end{lstlisting}
        \end{minipage}
    
        \caption{\textbf{Prettier formatting effects.} File: \texttt{index.js} in repo \texttt{chaijs/pathval} at commit 7859e0e.}
        \label{fig:normalization_example_1}
        \addcontentsline{toc}{subsection}{Figure S\arabic{figure}: Prettier formatting effects}
    \end{figure}

    \newpage

    \subsection*{Example Dataset Entries}

    \begin{figure}[H]
        \centering
        \begin{minipage}{\linewidth}
            \textbf{Data Entry: Positive Sample (Vulnerable)}
            \begin{lstlisting}[language=diff]
{
    "id": "leeoniya/uplot::b0fd072ea34b845be434841e42bf795ab840e210::src/utils.js::416::7",
    "body": "function assign(targ) {\r\n    const args = arguments;\r\n\r\n    for (let i = 1; i < args.length; i++) {\r\n        const src = args[i];\r\n\r\n        for (const key in src) {\r\n            if (isObj(targ[key]))\r\n                assign(targ[key], copy(src[key]));\r\n            else\r\n                targ[key] = copy(src[key]);\r\n        }\r\n    }\r\n\r\n    return targ;\r\n}",
    "label": 1,
    "paired_id": "leeoniya/uplot::5756e3e9b91270b303157e14bd0174311047d983::src/utils.js::420::7",
    "project": "leeoniya/uplot",
    "sha": "b0fd072ea34b845be434841e42bf795ab840e210",
    "file": "src/utils.js",
    "name": "assign",
    "loc": {
        "start_line": 416,
        "start_column": 7,
        "end_line": 429,
        "end_column": 1
    },
    "cwe": ["CWE-1321"],
    "cve": ["CVE-2024-21489"],
    "ghsa": ["GHSA-34q8-jcq6-mc37"],
    "snyk": ["SNYK-JS-UPLOT-6209224"],
    "other": ["RHSA-2024:8083"],
    "publish_time": {
        "day": 28,
        "month": 1,
        "year": 2024
    }
}\end{lstlisting}
        \end{minipage}
    
        \vspace{0.5cm}

        \begin{minipage}{\linewidth}
            \textbf{Data Entry: Negative Sample (Fixed)}
            \begin{lstlisting}[language=diff]
{
    "id": "leeoniya/uplot::5756e3e9b91270b303157e14bd0174311047d983::src/utils.js::420::7",
    "body": "function assign(targ) {\r\n    const args = arguments;\r\n\r\n    for (let i = 1; i < args.length; i++) {\r\n        const src = args[i];\r\n\r\n        for (const key in src) {\r\n            if (key != __proto__) {\r\n                if (isObj(targ[key]))\r\n                    assign(targ[key], copy(src[key]));\r\n                else\r\n                    targ[key] = copy(src[key]);\r\n            }\r\n        }\r\n    }\r\n\r\n    return targ;\r\n}",
    "label": 0,
    "paired_id": "leeoniya/uplot::b0fd072ea34b845be434841e42bf795ab840e210::src/utils.js::416::7",
    "project": "leeoniya/uplot",
    "sha": "5756e3e9b91270b303157e14bd0174311047d983",
    "file": "src/utils.js",
    "name": "assign",
    "loc": {
        "start_line": 420,
        "start_column": 7,
        "end_line": 435,
        "end_column": 1
    },
    "cwe": ["CWE-1321"],
    "cve": ["CVE-2024-21489"],
    "ghsa": ["GHSA-34q8-jcq6-mc37"],
    "snyk": ["SNYK-JS-UPLOT-6209224"],
    "other": ["RHSA-2024:8083"],
    "publish_time": {
        "day": 28,
        "month": 1,
        "year": 2024
    }
}\end{lstlisting}
        \end{minipage}
    
        \caption{\textbf{Dataset JSON format examples.} This figure illustrates a paired entry representing a Prototype Pollution vulnerability (CVE-2024-21489) and its corresponding patch. Each record contains an id, the function source code (\texttt{body}), vulnerability label, and metadata.}
        \label{fig:json_example}
        \addcontentsline{toc}{subsection}{Figure S\arabic{figure}: Dataset JSON format examples}
    \end{figure}

\end{document}